\documentclass[%
 reprint,
 amsmath,amssymb,
 aps,
]{revtex4-1}

\usepackage{graphicx}% Include figure files
\usepackage{dcolumn}% Align table columns on decimal point
\usepackage{bm}% bold math
\usepackage{lipsum}
\usepackage{tabularx}
\usepackage{array}
\usepackage{mathrsfs}
\usepackage{amssymb}
\usepackage{amsmath}
\usepackage{cases}  % To number the subequations of an equation
\usepackage{graphicx}% Include figure files
\usepackage{subfigure} %To insert multiple figures together
\usepackage{dcolumn}% Align table columns on decimal point
\usepackage{bm}% bold math
\usepackage{verbatim} % used for commenting on multiple paragraphs
\usepackage{graphicx}  
\usepackage{epstopdf}  
\usepackage{colortbl}
\usepackage{hyperref}

\begin{document}

\preprint{APS/123-QED}

\title{Hacking the quantum key distribution system by exploiting the avalanche transition region of single photon detectors}% Force line breaks with \\

\author{Yong-Jun Qian${}^{1,2,3,4}$}
\author{De-Yong He${}^{1,2,3,4}$}
\author{Shuang Wang${}^{1,2,3}$}
 \email{wshuang@ustc.edu.cn}

 \author{Wei Chen${}^{1,2,3}$}
 \author{Zhen-Qiang Yin${}^{1,2,3}$}
 \author{Guang-Can Guo${}^{1,2,3}$}
 \author{Zheng-Fu Han${}^{1,2,3}$}

\affiliation{${}^{1}$  CAS Key Laboratory of Quantum Information, University of Science and Technology of China, Hefei 230026, China\\ ${}^{2}$ CAS Center for Excellence in Quantum Information and Quantum Physics, University of Science and Technology of China, Hefei 230026, China\\ ${}^{3}$State Key Laboratory of Cryptology, P. O. Box 5159, Beijing 100878, China  \\ ${}^{4}$ These authors contributed equally to this work}

\date{\today}% It is always \today, today,

\begin{abstract}
Avalanche photodiode based single photon detectors, as crucial and practical components, are widely used in quantum key distribution (QKD) systems. For effective detection, most of these SPDs are operated in the gated mode, in which the gate is added to obtain high avalanche gain, and is removed to quench the avalanche. The avalanche transition region (ATR) is a certain existence in the process of adding and removing the gate. We first experimentally investigate the characteristic of the ATR, including in the commercial SPD and high-speed SPD, and then propose an ATR attack to control the detector. In the experiment of hacking the plug-and-play QKD system, Eve only introduces less than 0.5 \% quantum bit error rate, and almost leaves no traces of her presence including the photocurrent and afterpulse probability. We finally give possible countermeasures against this attack.

\begin{description}
\item[PACS numbers] 03.67.Dd
%You may use the \texttt{description} environment to structure your abstract;use the optional argument of the \verb+\item+ command to give the category of each item. 
\end{description}
\end{abstract}

\pacs{Valid PACS appear here}% PACS, the Physics and Astronomy
\maketitle

%\tableofcontents

\section{\label{sec:level1}Introduction}
Quantum key distribution (QKD) enables two remote parties, commonly known as Alice and Bob,  to share a string of secure keys \cite{bennet1984}. The eavesdropper, Eve, can not obtain any information without introducing errors. As the first application of quantum information  in wide fields,  the unconditional security of the BB84 QKD protocol has been proven \cite{Gisin2002,lo1999,Scarani2009,Gottesman2004,Advance2019}. 

 Generally, the source, modulators, and detectors are the main components in a QKD system. The models of these components are ideal in security proof, while these components have imperfections in real  life. The gaps between ideal models and imperfect components would leave loopholes for Eve,  and threaten the practical security  \cite{Brassard2000,WangXB2007,Huttner,Lutkenhaus,Fung2007,Xu2010,
Sun2011,Li2011,Zhao2008,Makarov2006,Makarov2016,Jain2011,	Lamas2007,Weier2011,Lydersen2010,YuanZL2010,Lydersenreply2010,
Gerhardt2011,LydersenThermal2010,Wiechers2011,Lydersensuperlinear2011,
Bugge2014,Jiang2013}. For instance, the original BB84 protocol requires a perfect single photon source, but most practical QKD systems employed the weak coherent source. The existence of multi-photon states give Eve a chink to implement the photon number splitting  attack, 
 until the decoy-state method was proposed \cite{Huttner,Lutkenhaus,Hwang2003,Wang2005,lo2005}.

Single photon detector (SPD) is an indispensable component for a BB84 QKD system, but as a complex one at the receiver's part, there are many loopholes that Eve can exploit to hack the system \cite{Zhao2008,Makarov2006,Makarov2016,Jain2011,	Lamas2007,Weier2011,Lydersen2010,YuanZL2010,Lydersenreply2010,
Gerhardt2011,LydersenThermal2010,Wiechers2011,Lydersensuperlinear2011,
Bugge2014,Jiang2013}. Recently, a series of hacking, named as detector control attacks, have been proposed  \cite{Lydersen2010,YuanZL2010,Lydersenreply2010,
Gerhardt2011,LydersenThermal2010,Wiechers2011,Lydersensuperlinear2011,
Bugge2014,Jiang2013}. One of the most famous attacks is detector blinding attack \cite{Lydersen2010,YuanZL2010,Lydersenreply2010,
Gerhardt2011,LydersenThermal2010}, the eavesdropper blinds the SPDs, and then remotely controls them to steal all keys without increasing the quantum bit error rate (QBER), but it requires bright illumination. Moreover, faint trigger pulses can also be used to control SPDs without extra blinding light, but the increased QBER is relatively high ($>$12\%) to be discovered \cite{Lydersensuperlinear2011}. 

Several strategies are proposed to defense these detector control attacks. The most attractive approach is measurement device independent protocol \cite{mdi2012,Pirandolamdi2012}, it can remove all detector side channel attacks, but  challenges of experiment and relative lower secure key rates need to be solved before commercial application.  Another approach is improving the existing systems, such as monitoring optical illumination \cite{Lydersen2010,Wang2014}, photocurrent \cite{YuanZL2011,	Lydersencomment2011,YuanZL2011reponse2011}, and afterpulse    \cite{Silva2012}, or randomly removing gates to check the clicks \cite{Wiechers2011}.  The improving approach  minimizes changes on the original system. But some of these methods  do not close all underlying loopholes or have not strictly  theoretical proof, new proposed attacks  may defeat the QKD system.

Here we propose an avalanche transition region (ATR) attack on  gated-mode avalanche photodiode (APD) detectors, which is widely used  in QKD systems \cite{Eisaman2011}. When the gate is on, the detector is in Geiger mode and sensitive to the single photon.  When the gate is off, it is in linear mode and can not detect the single photon. The ATR refers to the  region transitioned from Geiger mode to linear mode. When weak multiphoton signals   arrive at this region, the probability of being detected depends on the delay in the ATR and incident flux at matching basis and mismatching basis, Eve can implement ATR attack to steal all keys but introduce almost no errors ($ < 0.5 \% $). Moreover, since the incident flux is faint and avalanche gain factor in ATR is relatively small,  the strategies of monitoring illumination, photocurrent and afterpulse    are inoperative.

\section{ATR attack model}
\label{model}

\begin{figure}[b]
\resizebox{6 cm}{4cm}{
\includegraphics{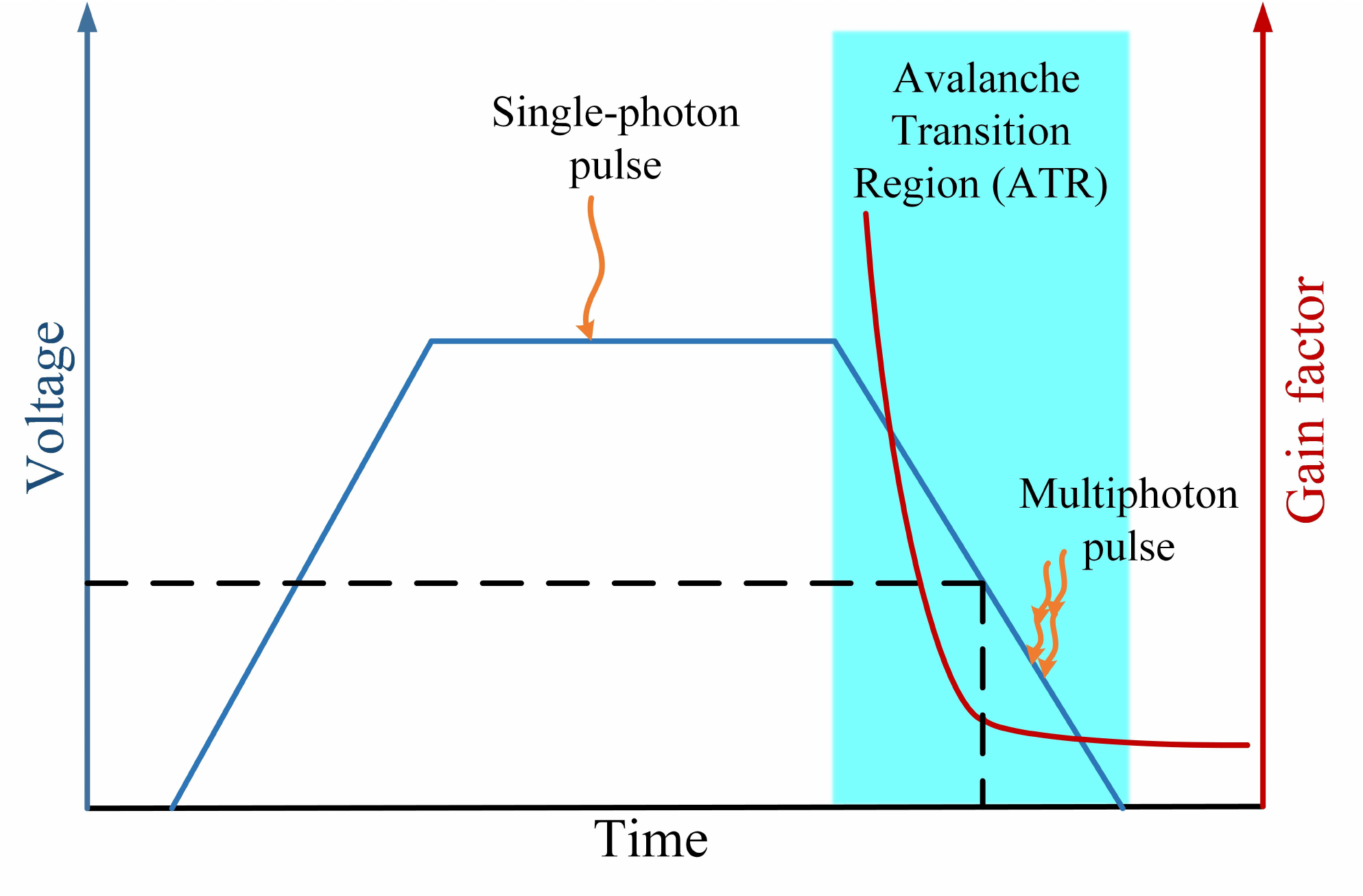}}% Here is how to import EPS art
\caption{\label{fig:ATR} Conceptual avalanche transition region.  $ V_{BD} $ denotes the breakdown voltage.  }
\end{figure}

For an  APD detector operated in the Geiger mode, a reverse bias voltage should be above the breakdown voltage ($V_{BD} $) when the gate is on. As is shown in Fig. \ref{fig:ATR},  in normal operation, a single-photon pulse arrives in the gate and creates a detectable macrosopic current signal. Then the gate should be removed to quench the avalanche \cite{Eisaman2011}. In reality,  there must be a voltage transition  in the process of removing the gate. Since the avalanche gain factor decreases with reduction of the bias voltage on APD \cite{Cova2004,Hyun1997,Ng2005}, this voltage transition region is also an avalanche transition region  (ATR). In this region, a single-photon pulse cannot be detected, but a multiphoton pulse (hundreds to thousands photons) would create a superimposed avalanche signals, whose amplitude is comparable with the one of a single avalanche signal created by a single-photon pulse in the gate.  

\begin{figure}[hbt]
\resizebox{8cm}{9.6cm}{
\includegraphics{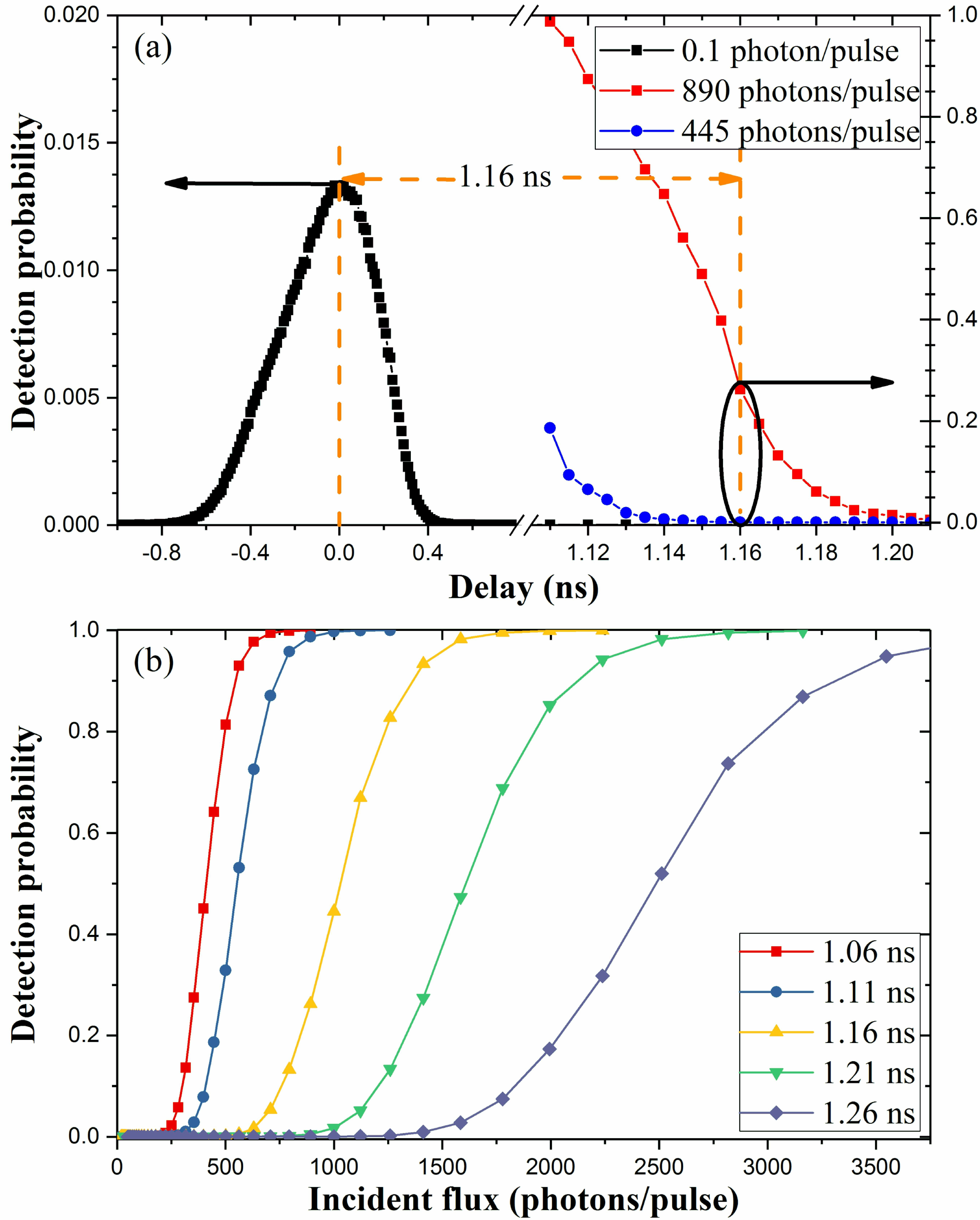}}% Here is how to import EPS art
\caption{\label{fig:behavior} Characteristic of ATR.	 (a) The detection probability versus delay position with different incident fluxes.   (b) The detection probability versus incident flux of the pulse at different delay positions.}
\end{figure}

The ATR of the APD detector is experimentally characterized by measurement of detection probability at different positions and with different incident fluxes. The commercial  SPD (id201, id Quantique) under test is based on the InGaAs APD and set 2.5 ns gate width. Under 1 MHz gating rate, this SPD is illuminated by a variable intensity pulse with 1 MHz repetition rate, and 30 ps temporal width at 1550 nm. As shown in Fig. \ref{fig:behavior}, the detection probability of this SPD changes a lot with the position and incident flux. When the incident flux is 0.1 photon/pulse (the black curve in Fig. \ref{fig:behavior}(a)), the delay between the incident pulse and gate signal is carefully tuned to get maximum detection probability. The maximum detection probability is about 1.33 \% (corresponding to 13.4 \% detection efficiency) during the gate signal region, and the corresponding delay time is set  as the zero point.  To show the characteristics of the ATR clearly, we only display the tested data in the range from the position of  1.06 ns delay to 1.26 ns delay in Fig. \ref{fig:behavior}. When the incident flux increases to hundreds of photons level, the corresponding detection probability  increases a lot even in the ATR, but reduces as the position away from the gate signal. In the region from 1.11 ns to 1.21 ns, the reducing ranges of the detection probability with different incident fluxes  are different, for instance, the range is from 98.7 \% to 0.52 \% for   890 photons/pulse flux (red curve of Fig. \ref{fig:behavior}(a)), from 18.6 \% to 0.04 \%  for 445 photons/pulse flux (blue curve of Fig. \ref{fig:behavior}(a)), and remains at the dark count probability level for 0.1 photons/pulse flux (black curve of Fig. \ref{fig:behavior}(a)).  Furthermore, the detection probability is also tested with increase of incident flux for five given positions in the ATR (their delay values are 1.06 ns, 1.11 ns, 1.16 ns, 1.21 ns, 1.26 ns, respectively). For each curves in  Fig. \ref{fig:behavior}(b), the detection probability starts rising gently, then increasing steeply, and  finally slowing down to saturation. It is obvious that the steepness of the increase of detection probability is weakened as the position in the ATR is far away from the gate signal. The farther in the ATR the position is, the lower the bias voltage on APD is, the smaller the avalanche gain factor is, then more incident photons are needed to create a superimposed signal with the same amplitude.

The property that the detection probability increases steeply  with the incident flux in the ATR would be used by Eve to control the APD detector operated in the gated mode. Taking the position of 1.16 ns delay for example, the detection probability is 26.2 \% when the incident flux is 890 photons/pulse, and reduces to 0.083 \% when the incident flux halves to 445 photons/pulse. Thus in order to control Bob's detector, Eve could send encoded multiphoton pulses with 890 photons/pulse flux to Bob (assume Bob's decoding components are lossless), and make these pulses arrive at the detector with the position of 1.16 ns  delay. If Eve and Bob select matching bases, these encoded multiphoton pulses have high probability to be detected. If Eve and Bob select opposite bases, half flux of these encoded multiphoton pulses hit the detector, and cause very low probability to be detected. After Bob publicly acknowledges his detected signals, Eve would have almost identical bases choices and bit values with Bob. That means Eve could steal almost all information of the secure keys shared between Alice and Bob, but leave almost no trace if she employed such an attack, which is named as ATR attack.

ATR attack is a general challenge for QKD systems with APD detectors operated in the gated-mode, in which the ATR is a necessary existence. Except for the widely-used commercial InGaAs SPD, two types of the homemade SPD have also tested to characterize the ATR. One is similar to id201, operating at 1 MHz, the other is using the sine-wave filtering method, operating at 1 GHz \cite{HeDY2017}. The same as above, the delay with the maximum detection probability at 0.1 photons/pulse is set as the zero point. For 1 MHz homemade SPD, at the delay of 1.09 ns, the detection probability is 21.5 \% when the incident flux is 890 photons/pulse, and reduces to 0.107 \% when the incident flux halves to 445 photons/pulse. When the incident flux is 1000 photons/pulse, the detection probability is 36.3 \%, and reduces to 0.29 \% when the incident flux halves to 500 photons/pulse. For 1 GHz homemade SPD, at the delay of 300 ps, the detection probability is 40 \% when the incident flux is 1000 photons/pulse, and reduces to 1.53 \% when the incident flux halves to 500 photons/pulse.

\section{ attacking experiment on the QKD system}
\label{experiment}
\begin{figure}[hbt]
\resizebox{8cm}{5.5cm}{
\includegraphics{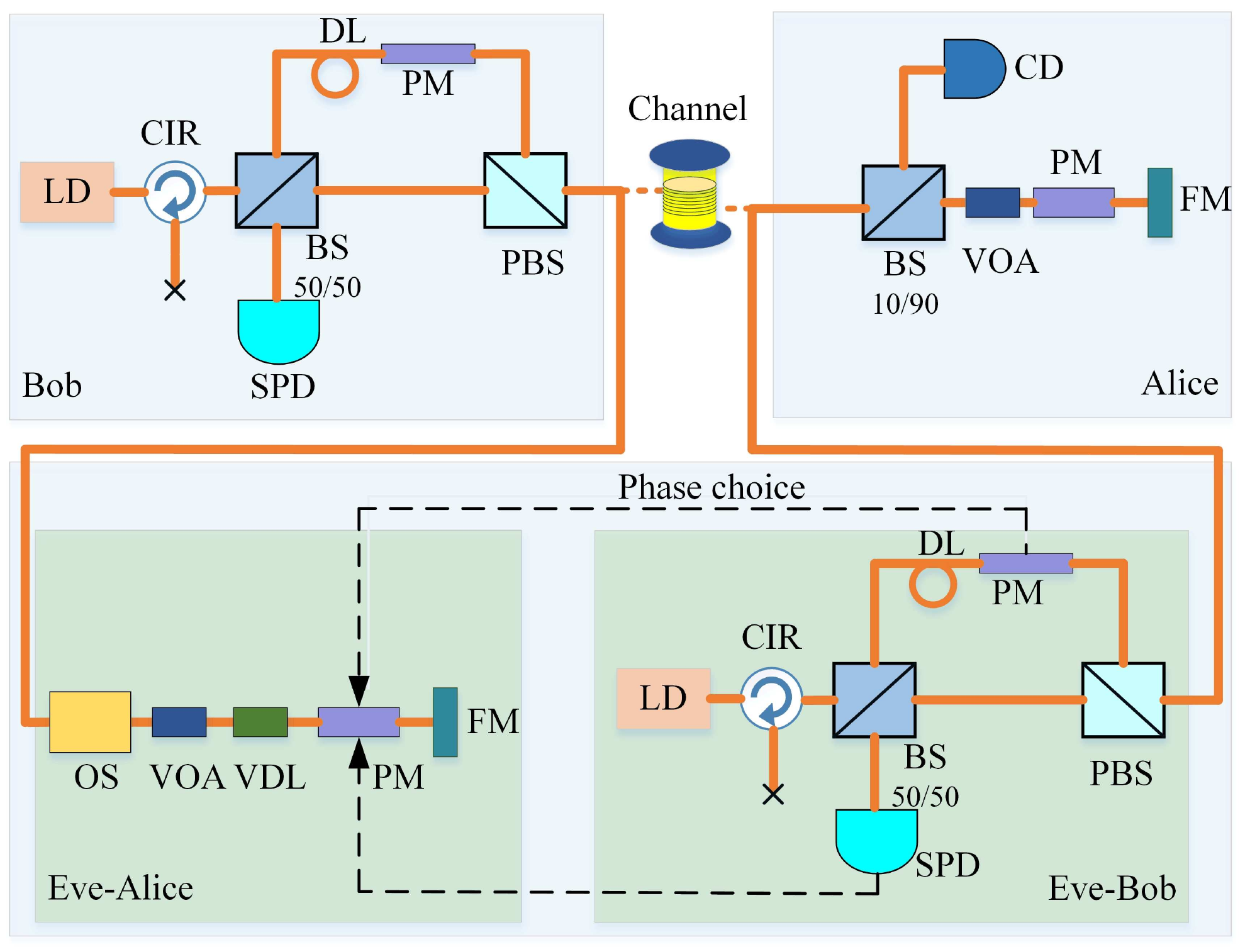}}% Here is how to import EPS art
\caption{\label{fig:experiment} Experimental demonstration of ATR attack on the plug-and-play QKD system. The original system is composed of Alice and Bob (upper part). And Eve's setup consists of Eve-Alice and Eve-Bob (bottom part). LD, laser diode; CIR, circulator;  BS, beam splitter; DL, delay line; PM, phase modulator; PBS, polarization beam splitter;  CD, classical detector; VOA, variable optical attenuator; VDL, variable delay line; OS, optical switch; FM, Faraday mirror.}
\end{figure}

The ATR attacking experiment is  demonstrated on the plug-and-play QKD system \cite{Stucki2002}, similar to the commercial QKD system id3110 Clavis2. The upper part of Fig. \ref{fig:experiment} shows the schematic setup of the system working in the normal operation, and BB84 protocol is implemented. To be simple and precise, four phases $\lbrace$ 0, $\frac{\pi}{2}$, $\pi$, $\frac{3\pi}{2}$ $\rbrace$  are randomly modulated at Bob's site, thus only one SPD is needed to implement the whole BB84 protocol. The above measured id201 is used to detect the output signal of the 50/50 beam splitter (BS) at Bob's site.

\begin{figure}[hbt]
\resizebox{8cm}{5cm}{
\includegraphics{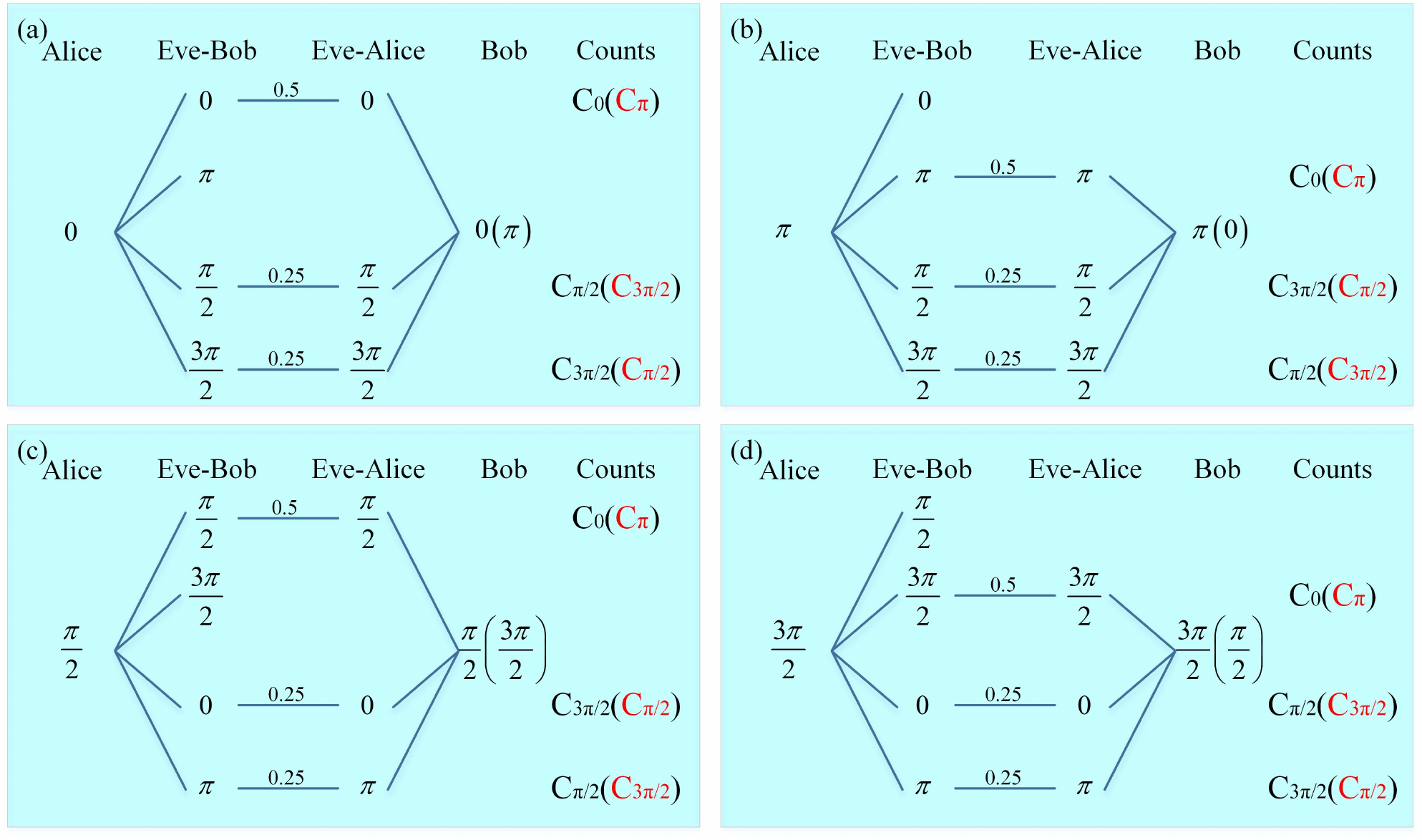}}% Here is how to import EPS art
\caption{\label{fig:tree} Detection tree of ATR attack combined with an intercept-resend strategy. In subfigures, each column indicates the phase the one chooses. The detection counts marked by red colour indicate error counts introduced by Eve.}
\end{figure}
 
Combing with an intercept-resend strategy, Eve can steal Alice and Bob's secret keys without being discovered. Eve's setup is shown in the bottom part of Fig. \ref{fig:experiment}, which consists of Eve-Bob and Eve-Alice. First, Eve-Bob sends strong pulses to Alice, and randomly chooses one of  $\lbrace$ 0, $\frac{\pi}{2}$, $\pi$, $\frac{3\pi}{2}$ $\rbrace$  phases after receiving the coding signals from Alice. From the responses of the SPD and the chosen phases, Eve can guess the phases that Alice encodes on the pulses, though the possibility of correct guess is 50 \%. Then, Eve-Alice intercepts the strong pulses from Bob, and modulates the guessed phases on these pulses, and resends these multiphoton pulses to Bob. The key of the ATR attack is that Eve-Alice should carefully control the delay of these coding pulses to make them arrive at the ATR of Bob's SPD, and also the incident flux of these pulses based on the characteristic of the ATR of SPD. The variable delay line (VDL) and variable optical attenuator (VOA) in Fig. \ref{fig:experiment} are used to tune the delay and incident flux of the coding pulses. And the optical switch (OS) is used to control the resending number of multiphoton pulses.  As shown in the detection tree of the ATR attack (see Fig. \ref{fig:tree}), the pulses intercepted by Eve-Alice have  very high (or zero) probability to be detected by Bob's SPD if the phase difference between Eve-Alice and Bob is 0 (or  $\pi$), and the counts of Bob's SPD are denoted as $C_{0}$ and $C_{\pi}$, respectively, when Eve and Bob choose the matching bases; the pulses intercepted by Eve-Alice have  very low probability to be detected by Bob's SPD if the phase difference between Eve-Alice and Bob is $\frac{\pi}{2}$ or $\frac{3\pi}{2}$, and these counts are denoted as $C_{\pi/2}$ and $C_{3\pi/2}$, respectively, when Eve and Bob choose the opposite bases. Thus QBER of the attacked QKD system can be given as

\begin{equation}
\label{eq:QBER}
QBER=\frac{C_{3\pi/2}+C_{\pi/2}+2C_{\pi}}{2(C_{0}+C_{\pi/2}+C_{\pi}+C_{3\pi/2})}.
\end{equation}

Note that since four phases are randomly modulated and only one SPD is used at Bob's site, and similar processes are also treated at Eve's site, $C_{\pi}$ corresponds to the error counts coming from apparatus imperfections \cite{YuanGHzQKD2008}. When Eve-Alice and Bob choose the matching bases and their phase difference is $\pi$, very small part of the multiphoton pulses would hit the SPD, and the detection probability of this small flux at ATR is very close to 0, so $ C_{\pi} $ is almost equal to 0. When Eve-Alice and Bob choose the opposite bases, we have $C_{\pi/2}=C_{3\pi/2}$. Suppose Eve-Alice resends $M$ multiphoton pulses per second, and the detection probability of Bob's SPD is denoted as $ P_{f} $ with full incident flux, and $ P_{h} $ with half incident flux, the counts can be expressed as $ C_{0}=\frac{1}{4}M \cdot P_{f} $, $C_{\pi/2}=C_{3\pi/2}= \frac{1}{4}M \cdot P_{h} $. Then Eq. \eqref{eq:QBER} can be simplified by

\begin{equation}
\label{eq:simQBER}
QBER\simeq\frac{C_{\pi/2}}{C_{0}+2C_{\pi/2}}=\frac{P_{h}}{P_{f}+2P_{h}}.
\end{equation}

\begin{figure}[hbt]
\resizebox{6cm}{5cm}{
\includegraphics{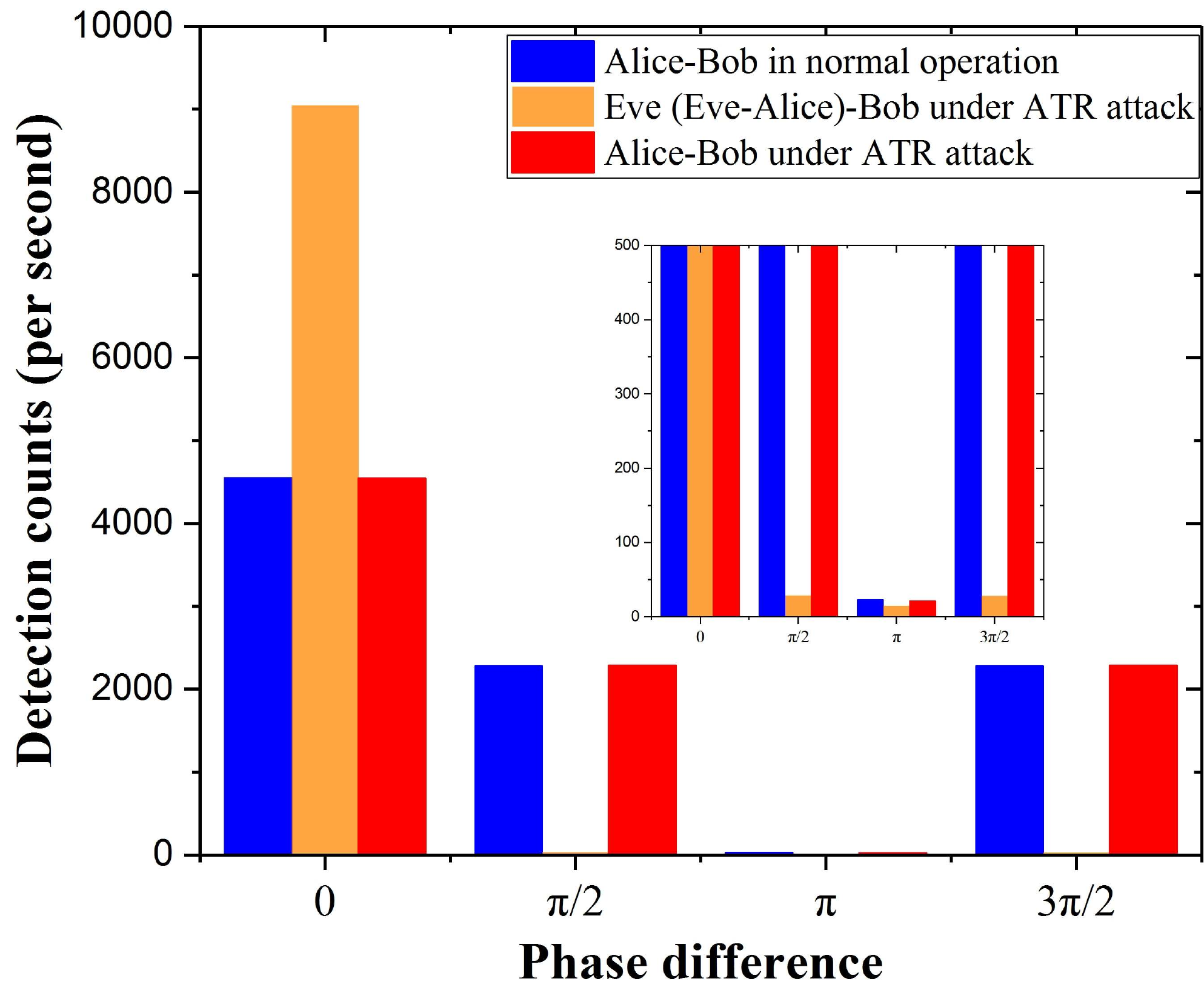}}% Here is how to import EPS art
\caption{\label{fig:result} Bob's detection counts versus phase difference under different perspectives. }
\end{figure}

In the ATR attacking experiment, Eve-Alice carefully tunes her VDL to make the resending multiphoton pulses arrive at the position of 1.16 ns delay, and tunes her VOA to make the flux hitting Bob's SPD be close to 890 photons/pulse if the phase difference between Eve-Alice and Bob is $ 0 $, and also controls her OS to ensure the counts of Bob's SPD are nearly unchanged. The experimental results are shown in Fig. \ref{fig:result}, which present Bob's detection counts versus phase difference under different perspectives. The first histogram (blue one) at each phase difference corresponds to the normal operation without Eve, the phase difference is between Alice and Bob. The second and third histograms correspond to the case with ATR attack, the phase difference in the second one (orange one) is between Eve (Eve-Alice) and Bob, and in the third one (red one) is between Alice and Bob. It is obvious that Eve has totally controlled Bob's SPD, and the statistical counts between Bob and Alice are almost identical before and after ATR attack. More importantly, QBER of the QKD system is nearly unchanged after being attacked. According to the second histogram (orange one) at each phase difference and Eq. \eqref{eq:QBER}, QBER under ATR attack is approximately 0.48 \%, which originates from imperfections of Eve's apparatus and  characteristic of ATR of Bob's SPD. At the position of 1.16 ns delay with 890 photon/pulse full incident flux, $ P_{f}=26.2 \%$, $ P_{h}=0.083 \% $,  if Eve's apparatus were perfect, the QBER introduced by ATR attack is approximately 0.31 \% according to Eq. \eqref{eq:simQBER}. Based on the ATR attack, Eve can obtain all information of the secure keys by hiding her presence in QBER of the QKD system, since the QBER introduced by this attack is small enough.

Here we want to emphasize that the goal of research on quantum hacking is to enhance the practical security of QKD, though the way of openly discovering and closing security loopholes \cite{Lydersen2010}, the ATR attack experiment is demonstrated in a proof-in-principle manner with only one SPD. Since the ATR attack is time-sensitive, the effectiveness of this attack would be reduced if two or more SPDs were used in the QKD system and they had different properties in the ATR. For example, with the incident flux of 890 photons/pulse, the chosen hacking position for the commercial SPD is at the delay of 1.16 ns, the QBER introduced by Eve is approximately 0.31 \%; while the chosen hacking position for the homemade SPD is at the delay of 1.09 ns, the QBER introduced by Eve is approximately 0.49 \%.  For these two SPDs, the hacking positions have a time mismatch of 0.07 ns. However, this mismatch could be compensated by exploiting the weakness of the calibration routine\cite{Jaincal2011}, which has been proposed by Jain et al. In reference \cite{Jaincal2011}, Eve causes a temporal separation up to 0.45 ns between two SPDs in a commercial QKD system.

\section{Countermeasures}

In the ATR attack, Eve resends the attacking pulses to the ATR of Bob's SPD, and totally controls the detector. Several countermeasures have been proposed to defeat the detector control attacks.  We first discuss some existing possible countermeasures against the ATR attack.

\begin{figure}[hbt]
\resizebox{7cm}{5cm}{
\includegraphics{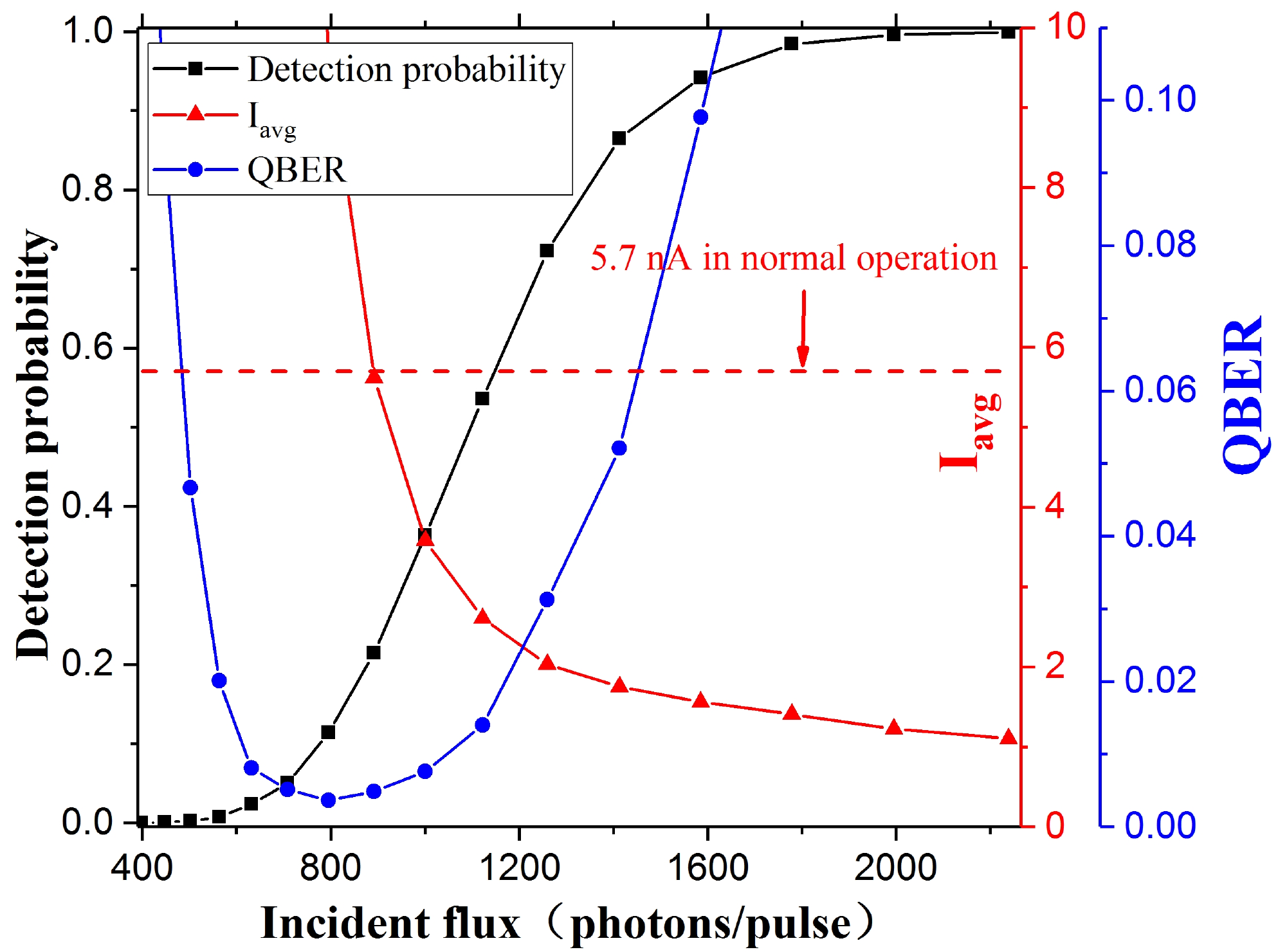}}% Here is how to import EPS art
\caption{\label{fig:current} The detection probability, average photocurrent ($ I_{avg} $) and  QBER versus incident flux. The detection probability is measured in the experiment. The average photocurrent is calculated with Eq. \eqref{photocurrent} based on the measured photocurrents per detection $ i_{f} $ and $ i_{h} $ and detection probabilities $ P_{f} $ and $ P_{h} $, in condition that Bob's detection count remains the same as the one in normal operation. QBER introduced by the ATR attack is calculated according to Eq. \eqref{eq:simQBER} based on the measured detection probabilities $P_{f} $ and $ P_{h} $. }
\end{figure}

Monitoring the photocurrent of the APD is  an effective way to detect most detector control attacks\cite{YuanZL2011,	 YuanZL2011reponse2011}, since these attacks would leave an obvious fingerprint of high photocurrent (more than 40 times stronger than the one under normal operation). But in ATR attack, the photocurrent could be equal to or even less than the one under normal operation. Suppose the photocurrent per detection count with full and half incident flux are $i_{f}$ and $ i_{h}$, respectively, which can be obtained when we measure the detection probability of Bob's SPD under full and half incident flux. Eve-Alice resends $M$ attacking pulses per second, and Bob gets $C$ detection counts, then $ C=C_{0}+C_{\pi/2}+C_{\pi}+C_{3\pi/2} \simeq \frac{1}{4}M(P_{f}+2P_{h})$. If Bob monitored the photocurrent of his SPD, the average photocurrent would be

\begin{equation}
\label{photocurrent}
\begin{aligned}
I_{avg} &=\frac{C(P_{f} \cdot i_{f}+2P_{h} \cdot i_{h})}{P_{f}+2P_{h}}
&,
\end{aligned}
\end{equation}
where the background photocurrent of SPD has been neglected. To measure the photocurrent characteristic under ATR attack, id201 was replaced by the similar homemade SPD, and the attacking position was changed from 1.16 ns to 1.09 ns delay. With different incident fluxes, the corresponding detection probability (black points), the average photocurrent (red points) and the QBER (blue points) are shown in Fig. \ref{fig:current}. The red dashed line denotes the average photocurrent under normal operation with the value of 5.7 nA, and the corresponding detection count is approximately $ 9.11 \times 10^{3} $ per second. When the incident flux is 890 photons/pulse, the photocurrent per detection count with full and half flux are 0.287 pA/count and 33.832 pA/count, respectively. The average photocurrent is about 5.6 nA, and QBER is less than 0.5 \%. From the point of view of the average photocurrent and QBER, it is hard to detect the ATR attack. When the incident flux is further increased, the average photocurrent would decrease in condition that the detection count keeps unchanged. Thus monitoring the average photocurrent is ineffective to detect the ATR attack.

In the second approach, afterpulses caused by macroscopic APD current are non-negligible \cite{YuanZL2011reponse2011}. For instance, the afterpulse probability increased from 1.79 \% to 76.6 \% after the after-gate attack \cite{ Silva2012}. For id201 in the QKD system under ATR attack in Sec. \ref{experiment}, the incident flux at the position of 1.16 ns are $  \lbrace 890, \ 445, \ 0, \ 445  \rbrace$  photons/pulse with equal probability, respectively, corresponding to the four phase differences $\lbrace$ 0, $\frac{\pi}{2}$, $\pi$, $\frac{3\pi}{2}$ $\rbrace$ between Eve-Alice and Bob. Following the method proposed by Yuan \textit{et al.} \cite{YuanZLafterpulse2007}, the afterpulse probability is measured with the value of 0.57 \%, which is consistent with the low QBER under ATR attack. So the second approach is also ineffective to find the ATR attack.

\begin{figure}[hbt]
\resizebox{8cm}{3.5cm}{
\includegraphics{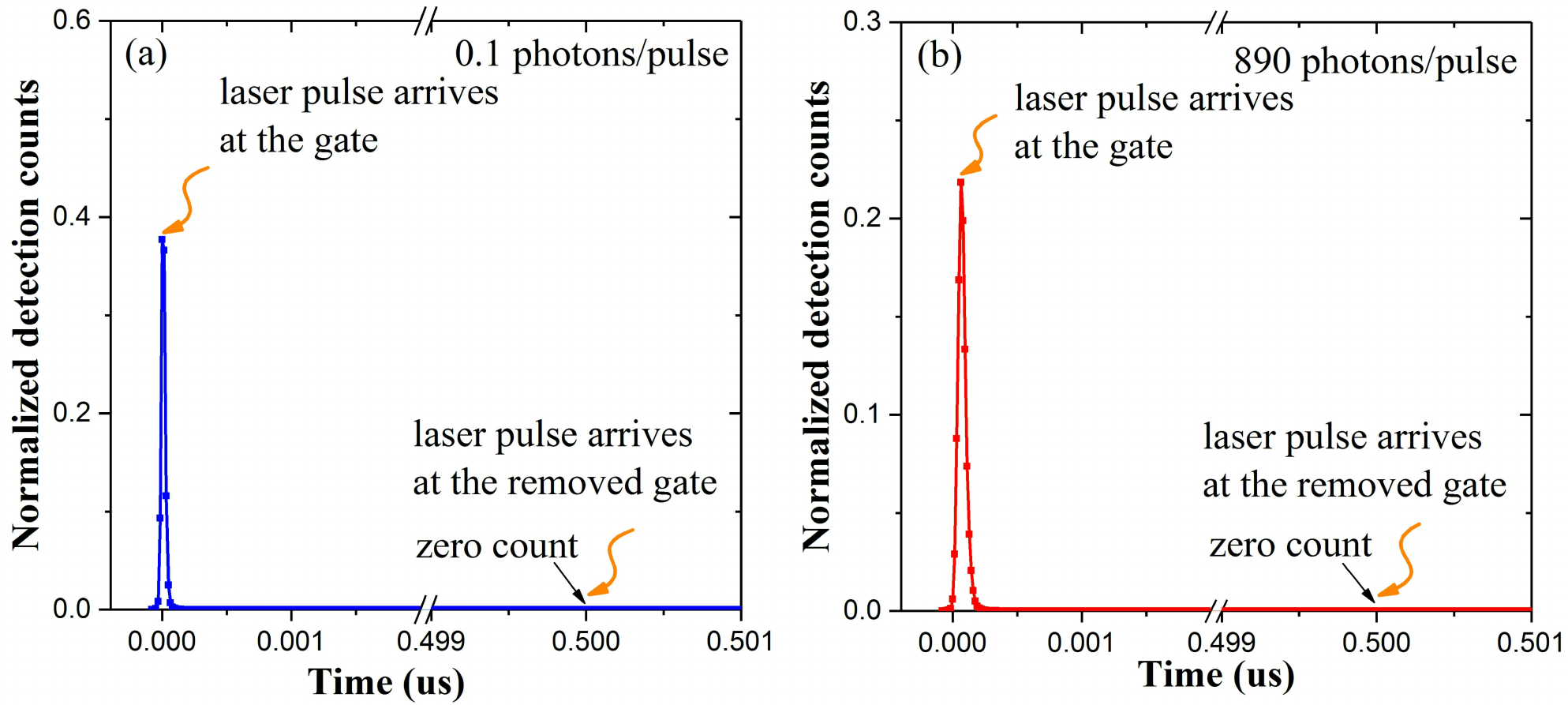}}% Here is how to import EPS art
\caption{\label{fig:remove}  Check detection counts at the gate and at the removed gate. The SPD is triggered with 1 MHz  rate and illuminated by a laser pulse train with 2 MHz  rate. (a)  The laser pulse train arrives at the zero point in the gate with 0.1 photons/pulse. (b) The laser pulse train  arrives at   1.16 ns delay with 890 photons/pulse. The timing resolution is 16 ps.  }
\end{figure}

A third possible countermeasure is randomly removing gates and checking clicks at the positions without gates \cite{Wiechers2011}, which is based on the property that clicks still occur under attacks. The effectiveness of this countermeasure against ATR attack is checked by the following experiment: the SPD id201 is triggered with 1 MHz repetition rate, and illuminated by a pulse train with 2 MHz rate, this is equivalent to remove half of 2 MHz gates. In normal case, the pulse train with 0.1 photons/pulse flux arrives at the zero point (with maximum detection probability). The temporal distribution of normalized detection counts is shown in Fig. \ref{fig:remove}(a), clicks only occur with gates, and there are no clicks at the removed gates. In ATR attack case, the pulse train with 890 photons/pulse flux arrives at the position of 1.16 ns delay. The corresponding temporal distribution of normalized detection counts is shown in  Fig. \ref{fig:remove}(b), which is similar with normal case, there are still no clicks at the removed gates. Since the output of the ATR attacking pulse mainly depends on the avalanche gain factor of APD, if the gate was removed, the gain factor would be very small, no clicks would occur. Hence removing gate is also invalid to detect the ATR attack.

\begin{figure}[hbt]
\centering
\resizebox{6cm}{4.5cm}{
\includegraphics{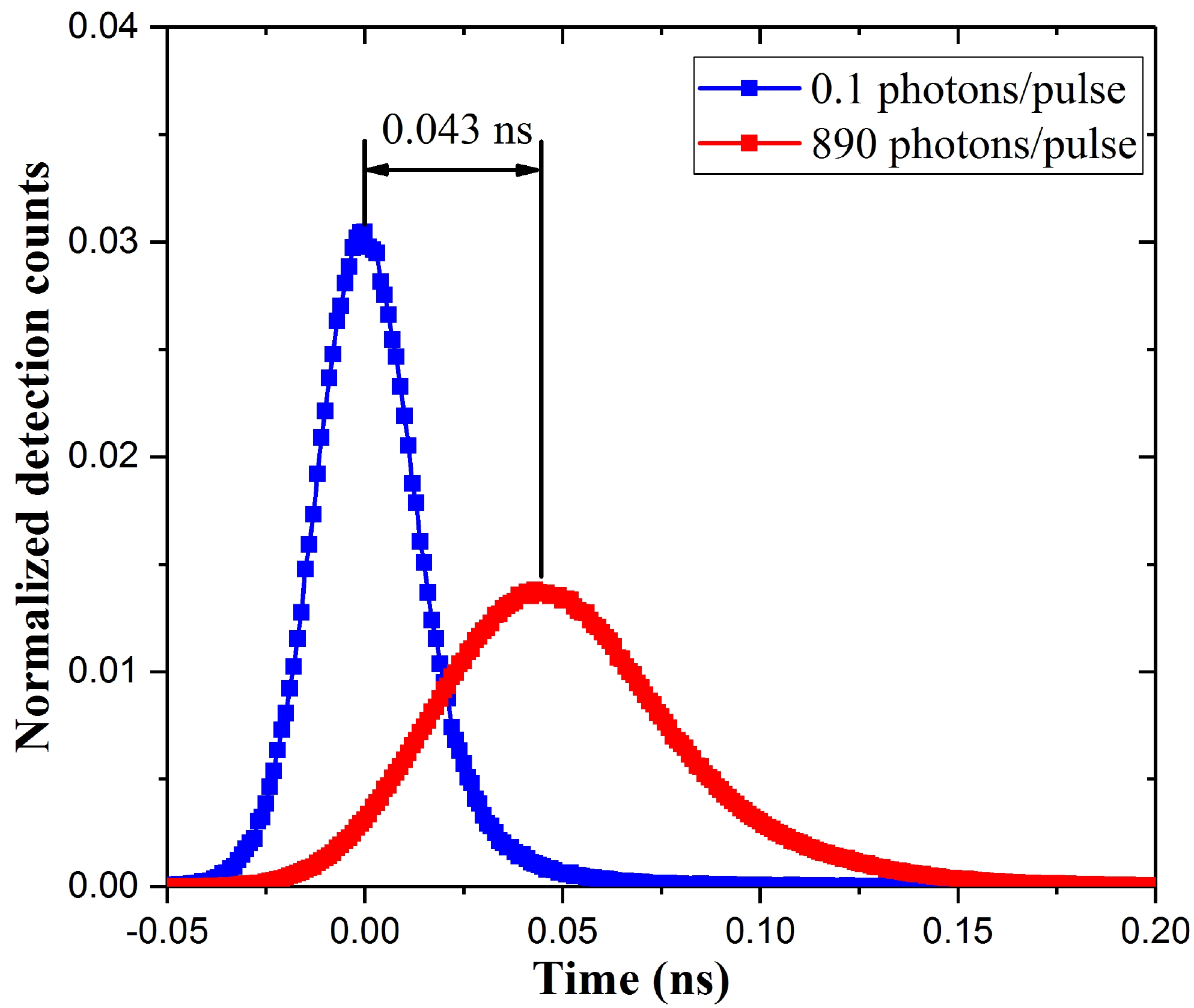}}% Here is how to import EPS art
\caption{\label{fig:histogram}  Temporal distribution of normalized detection counts. The timing resolution is 1 ps.}
\end{figure}

The three existing countermeasures cannot effectively detect the proposed ATR attack. However, we found that the temporal distribution of detection counts under ATR attack is different from the one under normal operation. The temporal distribution of detection counts of id201 is measured with 1 ps timing resolution, as shown in Fig. \ref{fig:histogram}. Under normal operation (blue one in Fig. \ref{fig:histogram}), the detector is illuminated by 0.1 photons/pulse incident flux at the zero point. Under ATR attack (red one in Fig. \ref{fig:histogram}), the detector is illuminated by 890 photons/pulse incident flux at 1.16 ns delay position. The temporal distribution of detection counts under normal operation is relatively concentrated in a small range (from -0.05 ns to 0.05 ns). While, the temporal distribution under ATR attack is relatively widespread (from -0.05 ns to 0.15 ns), and its center is delayed about 0.043 ns. These differences might be an approach to find ATR attack.

Additionally, since the avalanche gain factor in the ATR of APD is time-sensitive, QBER of the QKD system introduced by ATR attack would increase if the time jitter of the gate become large. For instance, when the setting of id201-`` Trigger delay " -is changed from `` Bypass "  to  `` Set 15 ns ", the full-width at half-maximum  of the time jitter of the gate (through measuring the `` Gate out " signal) increases from 19 ps to 65 ps. Then  at the position  of 1.16 ns delay, the detection probability with 890 photons/pulse incident flux increases from 26.2 \% to 97.6 \%, and the one with half flux (445 photons/pulse) increases from 0.083 \% to 44.9 \%, now the QBER introduced by ATR attack increases up to 24 \%. Still, Eve could change the incident flux to reduce her introduced QBER. If the full incident flux is changed to 400 photons/pulse, the detection probability is 31.3 \% for full flux, and 0.56 \% for half flux, the QBER introduced by ATR attack is about 1.7 \%.

\section{Conclusion}

In this paper, we propose and demonstrate an attack strategy exploiting the characteristic of ATR of the gated-mode SPD. The proposed ATR attack is a general challenge for QKD systems employing APD-based detectors operated in the gated-mode. Through choosing proper attacking position in ATR and incident flux of attacking pulses, Eve could almost completely control Bob's detector. In the attacking experiment on the plug-and-play QKD system, detection counts with different phase differences and QBER are identical before and after the ATR attack. Since the QBER introduced by Eve is less than 0.5 \%, Eve could hide her presence in the original error under the normal operation. And also, three existing countermeasures against detector control attacks are invalid to detect the proposed ATR attack. However, based on the experimental observations, we propose possible countermeasures to reveal the attack. The ATR attack highlights the importance of detection signals in practical QKD systems.

\begin{acknowledgments}
This work has been supported by the National Natural Science Foundation of China (Grant Nos. 61622506, 61575183, 61627820, 61475148, 61675189), the National Key Research And Development Program of China (Grant Nos.2018YFA0306400, 2016YFA0302600), Anhui Initiative in Quantum Information Technologies. 

Y-J. Q. and D-Y. H. contributed equally to this work.
\end{acknowledgments}

\nocite{*}


\begin{thebibliography}{40}
\bibitem{bennet1984} 
C. H. Bennet and G. Brassard, Quantum cryptography: public key distribution and coin tossing Int, in Proceedings of the IEEE International Conference on Computers, Systems and Signal Processing, Bangalore, India, 1984 (IEEE, New York, 1984), pp. 175-179.
\bibitem{Gisin2002}  N. Gisin, G. Ribordy, W. Tittel, and H. Zbinden, Quantum cryptography, Rev. Mod. Phys. 74, 145 (2002).
\bibitem{lo1999} H.-K. Lo and H. F. Chau, Unconditional security of quantum key distribution over arbitrarily long distances, Science 283, 2050 (1999).
\bibitem{Scarani2009} V. Scarani, H. Bechmann-Pasquinucci, N. J. Cerf, M. Du\v{s}ek, N. L\"{u}tkenhaus, and M. Peev, The security of practical quantum key distribution, Rev. Mod. Phys. 81, 1301 (2009).
\bibitem{Gottesman2004}D. Gottesman, H.-K. Lo, N. L\"{u}tkenhaus and J. Preskill, Security of quantum key distribution with imperfect devices, Quantum Inf. Comput. 4, 325 (2004).
\bibitem{Advance2019}S. Pirandola, U. L. Andersen, L. Banchi, M. Berta, D. Bunandar, R. Colbeck, D. Englund, T. Gehring, C. Lupo, C. Ottaviani, J. Pereira, M. Razavi, J. S. Shaari, M. Tomamichel, V. C. Usenko, G. Vallone, P. Villoresi and P. Wallden, Advances in Quantum Cryptography, arXiv:1906.01645 (2019).
\bibitem{Brassard2000}G. Brassard, N. L\"{u}tkenhaus, T. Mor, and B. C. Sanders, Limitations on practical quantum cryptography, Phys. Rev. Lett. 85, 1330 (2000).
\bibitem{WangXB2007}X. B. Wang, Decoy-state quantum key distribution with large random errors of light intensity, Phys. Rev. A 75, 052301 (2007).
\bibitem{Huttner}B. Huttner, N. Imoto, N. Gisin,  and T. Mor, Quantum cryptography with coherent states, Phys. Rev. A 51, 1863 (1995).
\bibitem{Lutkenhaus} N. L\"{u}tkenhaus and M. Jahma, Quantum key distribution with realistic states: photon-number statistics in the photon-number splitting attack, New J. Phys. 4, 44 (2002).
\bibitem{Fung2007} C. H. F. Fung, B. Qi, K. Tamaki,  and H. K. Lo, Phase-remapping attack in practical quantum-key-distribution systems, Phys. Rev. A 75, 032314 (2007).
\bibitem{Xu2010} F. Xu, B. Qi,  and H. K. Lo,  Experimental demonstration of phase-remapping attack in a practical quantum key distribution system, New J. Phys. 12, 113026 (2010).
\bibitem{Sun2011}S. H. Sun, M. S. Jiang, and L. M. Liang, Passive Faraday-mirror attack in a practical two-way quantum-key-distribution system, Phys. Rev. A 83, 062331 (2011).
\bibitem{Li2011} H. W. Li, S. Wang, J. Z. Huang, W. Chen, Z. Q. Yin, F. Y. Li, Z. Zhou, D. Liu, Y. Zhang, G. C. Guo, W. S. Bao, and Z. F. Han, Attacking a practical quantum-key-distribution system with wavelength-dependent beam-splitter and multiwavelength sources, Phys. Rev. A 84, 062308 (2011).
\bibitem{Zhao2008} Y. Zhao, C. H. F. Fung, B. Qi, C. Chen, and H. K. Lo, Quantum hacking: Experimental demonstration of time-shift attack against practical quantum-key-distribution systems, Phys. Rev. A 78, 042333 (2008).
\bibitem{Makarov2006}V. Makarov, A. Anisimov, and J. Skaar, Effects of detector efficiency mismatch on security of quantum cryptosystems, Phys. Rev. A  74, 022313 (2006).
\bibitem{Makarov2016}V. Makarov, J. P. Bourgoin, P. Chaiwongkhot, M. Gagn\'{e}, T. Jennewein, S. Kaiser, R. Kashyap, M. Legr\'{e}, C. Minshull, and S. Sajeed, Creation of backdoors in quantum communications via laser damage, Phys. Rev. A 94, 030302 (2016).
\bibitem{Jain2011} N. Jain, C. Wittmann, L. Lydersen, C. Wiechers, D. Elser, C. Marquardt, V. Makarov,  and G. Leuchs, Device calibration impacts security of quantum key distribution, Phys. Rev. Lett. 107, 110501 (2011).
\bibitem{Lamas2007}A. Lamas-Linares and C. Kurtsiefer, Breaking a quantum key distribution system through a timing side channel, Opt. Express 15, 9388 (2007).
\bibitem{Weier2011} H. Weier, H. Krauss, M. Rau, M. F\"{u}rst, S. Nauerth, and H. Weinfurter, Quantum eavesdropping without interception: an attack exploiting the dead time of single-photon detectors, New J. Phys. 13, 073024 (2011).
\bibitem{Lydersen2010} L. Lydersen, C. Wiechers, C. Wittmann, D. Elser, J. Skaar,  and V. Makarov, Hacking commercial quantum cryptography systems by tailored bright illumination, Nat. Photonics 4, 686 (2010).
\bibitem{YuanZL2010}Z. L. Yuan, J. F. Dynes, and A. J. Shields, Avoiding the blinding attack in QKD, Nat. Photonics 4, 800 (2010).
\bibitem{Lydersenreply2010}L. Lydersen, C. Wiechers, C. Wittmann, D. Elser, J. Skaar, and V. Makarov, Avoiding the blinding attack in QKD, Nat. Photonics 4, 801 (2010).
\bibitem{Gerhardt2011}I. Gerhardt, Q. Liu, A. A. Lamas-Linares, J. Skaar, C. Kurtsiefer,  and V. Makarov, Full-field implementation of a perfect eavesdropper on a quantum cryptography system,  Nat. Commun. 2, 349 (2011).
\bibitem{LydersenThermal2010} L. Lydersen, C. Wiechers, C. Wittmann, D. Elser, and J. Skaar, Thermal blinding of gated detectors in quantum cryptography, Opt. Express 18, 27938 (2010).
\bibitem{Wiechers2011}C. Wiechers, L. Lydersen, C. Wittmann, D. Elser, J. Skaar, C. Marquardt, V. Makarov, and G. Leuchs, After-gate attack on a quantum cryptosystem, New J. Phys. 13, 013043 (2011).
\bibitem{Lydersensuperlinear2011} L. Lydersen, N. Jain, C. Wittmann,  \O. Marøy, J. Skaar, C. Marquardt, V. Makarov,  and G. Leuchs, Superlinear threshold detectors in quantum cryptography, Phys. Rev. A 84, 032320 (2011).
\bibitem{Bugge2014} A. N. Bugge, S. Sauge, A. M. M. Ghazali, J. Skaar, L. Lydersen and V. Makarov, Laser damage helps the eavesdropper in quantum cryptography,  Phys. Rev. Lett. 112,  070503 (2014).
\bibitem{Jiang2013} M. S. Jiang, S. H. Sun, G. Z. Tang, X. C. Ma, C. Y. Li and  L. M. Liang, Intrinsic imperfection of self-differencing single-photon detectors harms the security of high-speed quantum cryptography systems, Phys. Rev. A 88, 062335 (2013).
\bibitem{YuanZL2011}  Z. L. Yuan, J. F. Dynes and A. J. Shields, Resilience of gated avalanche photodiodes against bright illumination attacks in quantum cryptography, Appl. Phys. Lett. 98, 231104 (2011).
\bibitem{Lydersencomment2011}L. Lydersen, V. Makarov and J. Skaar, Comment on `Resilience of gated avalanche photodiodes against bright illumination attacks in quantum cryptography', Appl. Phys. Lett. 99, 196101 (2011).
\bibitem{YuanZL2011reponse2011}Z. L. Yuan, J. F. Dynes and A. J. Shields, Response to ``Comment on `Resilience of gated avalanche photodiodes against bright illumination attacks in quantum cryptography' '',  Appl. Phys. Lett. 99, 196101 (2011).
\bibitem{Silva2012}T. F. da Silva, G. B. Xavier, G. P. Tempor\~ao, and J. P. von der Weid, Real-time monitoring of single-photon detectors against eavesdropping in quantum key distribution systems, Opt. Express 20, 18911 (2012).
\bibitem{Wang2014}  S. Wang, W. Chen, Z.-Q. Yin, H.-W. Li, D.-Y. He, Y.-H. Li, Z. Zhou, X.-T. Song, F.-Y. Li, D. Wang, H. Chen, Y.-G. Han, J.-Z. Huang, J.-F. Guo, P.-L. Hao, M. Li, C.-M. Zhang, D. Liu, W.-Y. Liang, C.-H. Miao, P. Wu, G.-C. Guo and Z.-F. Han, Field and long-term demonstration of a wide area quantum key distribution network, Opt. Express 22, 21739 (2014).
\bibitem{Hwang2003}W.-Y. Hwang, Quantum key distribution with high loss: toward global secure communication, Phys. Rev. Lett. 91, 057901 (2003).
\bibitem{Wang2005} X.-B. Wang, Beating the photon-number-splitting attack in practical quantum cryptography, Phys. Rev. Lett. 94, 230503 (2005).
\bibitem{lo2005}H.-K. Lo, X. Ma, and K. Chen,  Decoy state quantum key distribution, Phys. Rev. Lett. 94, 230504 (2005).
\bibitem{mdi2012}H.-K. Lo, M. Curty, and B. Qi, Measurement-device-independent quantum key distribution, Phys. Rev. Lett. 108, 130503 (2012).
\bibitem{Pirandolamdi2012} S. L. Braunstein and S. Pirandola, Side-channel-free quantum key distribution, Phys. Rev. Lett. 108, 130502 (2012).
\bibitem{Eisaman2011} M. D. Eisaman,  J. Fan, A. Migdall, and S. V. Polyakov, Invited review article: Single-photon sources and detectors, Rev. Sci. Instrum. 82, 071101 (2011).
\bibitem{Cova2004} S. Cova, M. Ghioni, A. Lotito, I. Rech, and F. Zappa, Evolution and prospects for single-photon avalanche diodes and quenching circuits, J. Mod. Opt. 51, 1267 (2004).
\bibitem{Hyun1997} K. S. Hyun, and C. Y. Park, Breakdown characteristics in InP/InGaAs avalanche photodiode with pin multiplication layer structure, J. Appl. Phys. 81, 974 (1997).
\bibitem{YuanGHzQKD2008} Z.  L. Yuan, A.  R. Dixon, J.  F. Dynes, A.  W. Sharpe, and A.  J. Shields, Gigahertz quantum key distribution with InGaAs avalanche photodiodes, Appl. Phys. Lett. 92, 201104 (2008).
\bibitem{Ng2005}  J. S. Ng, C. H. Tan, J. P. R. David , and  G. J. Rees,  Effect of impact ionization in the InGaAs absorber on excess noise of avalanche photodiodes,  IEEE J. Quantum Electron. 41, 1092 (2005).
\bibitem{Stucki2002}D. Stucki, N. Gisin, O. Guinnard, G. Robordy, and H. Zbinden, Quantum key distribution over 67 km with a plug\&play system, New J. Phys. 4, 41 (2002).
\bibitem{YuanZLafterpulse2007}Z. L. Yuan, B. E. Kardynal, A. W. Sharpe, and A. J. Shields, High speed single photon detection in the near infrared, Appl. Phys. Lett. 91, 041114 (2007).
\bibitem{HeDY2017}D.-Y. He, S. Wang, W. Chen, Z.-Q. Yin, Y.-J. Qian, Z. Zhou,G.-C. Guo, and Z.-F. Han, Sine-wave gating InGaAs/InP single photon detector with ultralow afterpulse,  Appl. Phys. Lett. 110, 111104 (2017).
\bibitem{Jaincal2011} N. Jain, C. Wittmann, L. Lydersen, C. Wiechers, D. Elser, C. Marquardt, V. Makarov, and G. Leuchs, Device calibration impacts security of quantum key distribution, Phys. Rev. Lett. 107, 110501 (2011).
\end{thebibliography}

\end{document}